\documentclass[compactaffiliation]{Interspeech}

\interspeechcameraready

\title{Pushing the Limits of Beam Search Decoding \\ for Transducer-based ASR models}

\author[affiliation={1}, equalcontribution]{Lilit}{Grigoryan}
\author[affiliation={1}, equalcontribution]{Vladimir}{Bataev}
\author[affiliation={1}]{Andrei}{Andrusenko}
\author[affiliation={2}]{Hainan}{Xu}
\author[affiliation={2}]{Vitaly}{Lavrukhin}
\author[affiliation={2}]{Boris}{Ginsburg}

\affiliation[]{}{NVIDIA}{Armenia}
\affiliation[]{}{NVIDIA}{USA}
\email{\{lgrigoryan, vbataev, aandrusenko, hainanx, vlavrukhin\}@nvidia.com}

\usepackage{comment}

\usepackage{amssymb}

\let\emptyset\varnothing
\usepackage[noend]{algpseudocode}
\usepackage{algorithm}
\makeatletter
\patchcmd{\ALG@doentity}{\item[]\nointerlineskip}{}{}{} 
\renewcommand{\ALG@beginalgorithmic}{\footnotesize}       
\algrenewcommand\algorithmicrequire{\textbf{Input:}}    
\algrenewcommand\algorithmicensure{\textbf{Output:}}    

\usepackage{tabularx}
\usepackage{float}
\usepackage{multirow} 
\usepackage[noabbrev]{cleveref}
\makeatother

\begin{document}

\maketitle

\begin{abstract}

Transducer models have emerged as a promising choice for end-to-end ASR systems, offering a balanced trade-off between recognition accuracy, streaming capabilities, and inference speed in greedy decoding. However, beam search significantly slows down Transducers due to repeated evaluations of key network components, limiting practical applications. This paper introduces a universal method to accelerate beam search for Transducers, enabling the implementation of two optimized algorithms: ALSD++ and AES++. The proposed method utilizes batch operations, a tree-based hypothesis structure, novel blank scoring for enhanced shallow fusion, and CUDA graph execution for efficient GPU inference. This narrows the speed gap between beam and greedy modes to only 10–20\% for the whole system, achieves 14–30\% relative improvement in WER compared to greedy decoding, and improves shallow fusion for low-resource up to 11\% compared to existing implementations. All the algorithms are open sourced.

\end{abstract}

\section{Introduction}

The Recurrent Neural Transducer (RNN-T or Transducer) architecture~\cite{graves_transducer} is widely employed in Automatic Speech Recognition (ASR) systems, as it provides a balanced trade-off between recognition accuracy and inference speed during greedy decoding. Moreover, Transducer-based models are the preferred choice for streaming ASR due to their causal decoders.

However, when switching to beam search decoding, Transducers slow down significantly due to multiple evaluations of the Joint and Prediction Network (Decoder) components. This fact greatly limits their usability in some scenarios, such as low/middle-resource languages and context-biasing tasks (for example, shallow fusion with an external language model (LM)~\cite{ShallowFusionEC} or word boosting with a context graph~\cite{Andrusenko2024FastCF}), where multiple alternative hypotheses must be processed using beam search decoding. In this case, the inference time increases by several times (up to 5-10x) compared to greedy decoding.

One intuitive approach to address this issue is to reduce the hypothesis search space through pruning, similar to conventional hybrid ASR models \cite{Nolden2011AcousticLF}. \cite{Jain2019RNNTFL} introduced state and expand beams to prune hypotheses within the standard beam search algorithm. In \cite{alsd}, a transition from time-synchronous to Alignment-Length Synchronous Decoding (ALSD) is proposed to cut down on Joint network evaluations and hypothesis expansions without sacrificing accuracy. \cite{maes} presented an Adaptive Expansion Search (AES) that limits the maximum number of expansions per time frame and leverages the batched hypothesis computations. In \cite{Keren2023ATB}, a token-wise segmentation was proposed for faster audio processing.
\cite{kang2023fastparallel} limits model emission to one token per frame and proposes modifying the loss function to achieve better accuracy in such a setup.

Beyond the beam search algorithm, the Transducer architecture itself can be modified. The stateless Transducer introduced in \cite{stateless} replaces the LSTM-based decoder with a lightweight embedding of the last \textit{n} predicted tokens. 
In \cite{Wang2022AcceleratingRT}, the use of blank predictions from an auxiliary CTC head reduces the number of RNN-T decoder calls. \cite{tdt} introduces the Token-and-Duration Transducer (TDT), which predicts both a token and the number of frames it covers, allowing frame-skipping. A cascaded encoder for Transducer models is also proposed in \cite{Mahadeokar2022StreamingPT} to accelerate beam search decoding in streaming ASR.
Despite their promise to reduce beam search time, their speeds still remain significantly lower than greedy decoding.

Fast greedy search methods for Transducers have further diminished the need for beam search decoding. The label-looping algorithm \cite{article:labellooping} combined with an optimized CUDA implementation \cite{Galvez2024SpeedOL} narrows the performance gap between greedy Transducer and CTC decoding to only 10–20\%, making slow beam search a significant bottleneck.

In this paper, we aim to revitalize beam search for Transducer models. Inspired by advances in greedy search, we propose a universal method for beam search acceleration that allowed us to obtain new fast algorithms for RNN-T and TDT inference: ALSD++ and AES++. The proposed method incorporates several key innovations: (1)~efficient batch processing; (2)~a new hypothesis data structure for rapid operations such as storage, extension, comparison, and pruning; (3)~a novel approach for scoring blanks during shallow fusion; (4)~CUDA graphs for fast GPU decoding.
Our method reduces the speed gap between beam search and highly-optimized greedy search to only 10–20\%. Beyond speed, we demonstrate the advantages of beam search decoding for context-biasing tasks with an external n-gram LM. By using a GPU-efficient LM, we enable late pruning for hypotheses, which
collectively contributes to a relative improvement of 14–30\% in Word Error Rate (WER). All proposed algorithms are available in the NeMo toolkit\footnote{\begin{scriptsize}\url{https://github.com/NVIDIA/NeMo/pull/12729}\end{scriptsize}}~\cite{article:nemo}.

The main contribution of our work can be formulated as:
\begin{itemize}
    \item A universal beam search acceleration method for GPU, based on a new tree-based hypothesis data structure.
    \item Novel blank scoring method paired with late pruning for enhanced shallow fusion with external LM.
    \item Novel, fully batched beam search decoding algorithms for Transducer models: AES++ and ALSD++  with LM shallow fusion and CUDA graphs support.
    \end{itemize}

\section{Methods}
A beam decoding iteration typically involves several key sub-steps. First, in the expansion step, current hypotheses within the beam are extended with one or more new tokens. Next, in the merging step, hypotheses with identical token sequences are combined. Finally, pruning is applied to retain the top \textit{N} highest-scoring hypotheses for the next step.

A straightforward beam search implementation represents each hypothesis as a separate object \cite{article:nemo, article:espnet}, but while intuitive, it is not suitable for batched operations, that are preferred on GPUs. Existing beam decoding approaches further limit performance by processing audio streams sequentially, underutilizing available GPU resources.

\subsection{Batched beam search for Transducer-based models}
To address this, we propose a batched beam decoding approach that enables the simultaneous processing of multiple audio streams and tends to maximize batched operations by structuring hypotheses to leverage parallelism.

\textbf{Audio- and beam- level batching:} We extend the \textsl{BatchedHyps} structure from batched greedy search \cite{article:labellooping} by introducing \textsl{BatchedBeamHyps}, a GPU-efficient structure for compactly storing a batch of beam hypotheses. To avoid extra allocations, we maintain exactly $\mathrm{Beam}$ hypotheses throughout whole decoding process, updating them with exactly $\mathrm{Beam}$ expansions at each step. Some hypotheses may generate multiple expansions while others can be discarded, so to preserve the total number of hypotheses, we replace discarded hypotheses with additional expansions.

\begin{figure}[t]
    \centering
    \includegraphics[width=0.85\linewidth]{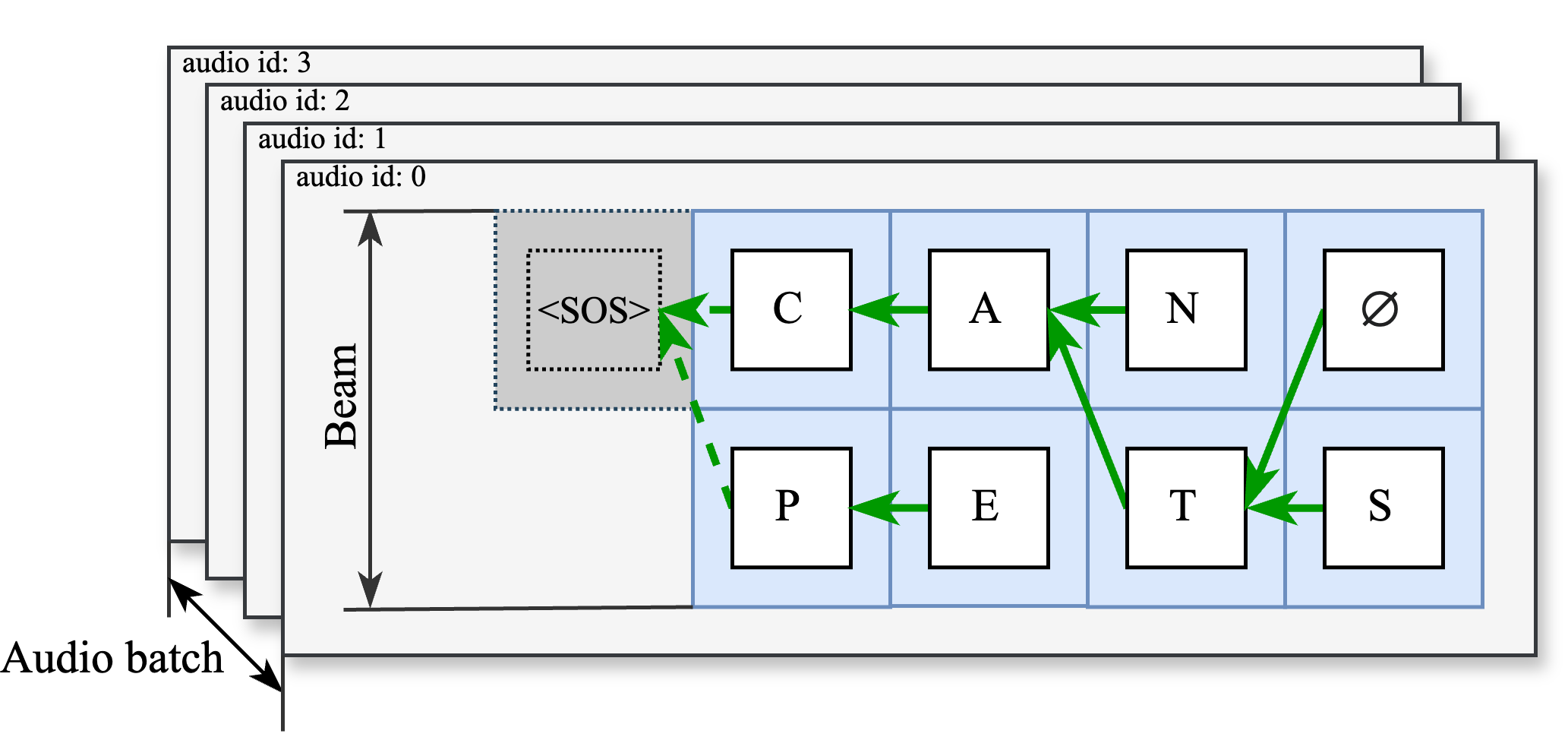}
    \vspace{-1em}
    \caption{Tree structure for batched hypotheses.}
    \label{fig:trie}
    \vspace{-2em}
\end{figure}

\textbf{Efficient hypothesis storing:} We implement a trie-like structure to store transcripts efficiently, enabling hypothesis expansion without duplicating entire transcripts (Fig. \ref{fig:trie}).

The tree is represented using two 3D tensors: $\mathrm{transcripts}$ (stores tokens for current step) and $\mathrm{transcripts\_ptrs}$ (stores backlinks to positions where preceding tokens are stored). Full transcript retrieval is performed in reverse order, from the latest token back to the first. For instance, for a hypothesis $\mathrm{hyp\_i}$ of length $\mathrm{u}$,  linked to an audio stream with index $\mathrm{audio\_id}$, the latest token is stored at $\mathrm{transcripts[audio\_id, hyp\_i, u]}$. The previous token is found at $\mathrm{transcripts[audio\_id, pref\_i, u-1]}$, where $\mathrm{pref\_i=transcripts\_ptrs[audio\_id, hyp\_i, u]}$ is the index of prefix at preceding step. This structure offers several advantages, particularly effective for large beam sizes and long sequences: 1) improves memory efficiency by eliminating the need to store common prefixes for multiple hypotheses, 2) it enables faster hypothesis expansion by simply adding a new token and updating the pointer structure, 3) reduces computational overhead by minimizing the amount of data that needs to be copied or moved during beam search

\textbf{Efficient hypothesis pruning and merging:} 
Due to flexible blank placements, alignments with different blank positions, but identical transcripts can be produced. E.g., sequences `\texttt{a$\emptyset$b}', `\texttt{a$\emptyset$$\emptyset$b}', `\texttt{ab$\emptyset$}' (where $\emptyset$ represents a blank symbol) all collapse to `\texttt{ab}'. This was first noted in the original RNN-T paper~\cite{graves_ctc} and is present in the mAES~\cite{maes} and ALSD~\cite{alsd} algorithms on which our work is based. 
While merging hypotheses improves the quality, comparing all hypotheses token by token results in linear complexity at each decoding step, which is costly for long sequences and large beam sizes, as it leads to quadratic complexity for the entire algorithm. To compare hypotheses in constant time (and thus make decoding complexity linear), we use hash-based transcript representations. Each hypothesis maintains an incremental hash, updated with each new non-blank token. Specifically, given an existing hash $H_t$ for a hypothesis of length $t$, the hash for the extended hypothesis $H_{t+1}$ is computed as:
\vspace{-2mm}
\begin{equation}
    H_{t+1}=(H_{t} \times P+T_{t+1}) \pmod M
\end{equation}

where $P$ is a chosen prime base, $M$ is a large prime modulus, and $T_{t+1}$ is the newly appended token. For sufficiently large $P$ and $M$, hash collisions are extremely rare (we observe less than 0.1\% in practice). To further reduce this risk, we compare transcript lengths and the latest tokens. With this approach, hypotheses can be compared in constant time, with roughly $\mathrm{Beam} \times \mathrm{Beam}$ integer comparisons. To maintain constant tensor sizes, instead of removing hypotheses, we replace a score with $-\infty$ to prevent their selection in future steps. Note that the same technique can be applied for prefix search \cite{graves_transducer}.

\textbf{CUDA Graphs:} During beam decoding, the GPU executes repeating small operations at each step, and the overhead from kernel launches can become substantial. Particularly, with smaller beam and batch sizes, each kernel performs fewer memory and computation tasks, making the kernel launch time potentially longer than the execution time itself. To mitigate this, we integrated CUDA Graphs into our decoding pipeline. This technique captures and replays a predefined sequence of GPU operations to reduce the overhead associated with kernel launches. As a result, we eliminate potentially redundant kernel launch costs and lower CPU-GPU synchronization overhead.

\begin{table*}[t!]
\centering
\caption{Table of decoding results with and without LM fusion. LM parameters are tuned on dev set across 4 fusion strategies. \textbf{Model:} RNN-T$({\small \sim} 1k, 24k)$, TDT. \textbf{Data:} SPGI, Europarl, SLURP test. \textbf{Strategy:} greedy, AES, ALSD and ALSD++ (ours), AES++ (ours). \textbf{Batch size:} 32. \textbf{Beam size:} 6. \textbf{CG}-denotes CUDA Graphs.}
\vspace{-1em} 
\resizebox{\textwidth}{!}{%
\begin{tabular}{lcc|rr|cc|rr|cc|rr}
\toprule
\multirow{2}{*}{\textbf{Decoding method}} & \multicolumn{4}{c|}{\textbf{SPGI}} & \multicolumn{4}{c|}{\textbf{Europarl}} & \multicolumn{4}{c}{\textbf{SLURP}} \\
\cmidrule{2-13}
& \multicolumn{2}{c|}{\textbf{WER} $\downarrow$} & \multicolumn{2}{c|}{\textbf{RTFx} $\uparrow$} & \multicolumn{2}{c|}{\textbf{WER} $\downarrow$} &  \multicolumn{2}{c|}{\textbf{RTFx} $\uparrow$} & \multicolumn{2}{c|}{\textbf{WER} $\downarrow$} & \multicolumn{2}{c}{\textbf{RTFx}$\uparrow$} \\
\midrule
\multicolumn{1}{l}{\textbf{RNN-T ($\sim$1k hours)}} 
& \multicolumn{1}{c}{no LM} & \multicolumn{1}{c|}{with LM} 
& \multicolumn{1}{c}{no LM} & \multicolumn{1}{c|}{with LM} 
& \multicolumn{1}{c}{no LM} & \multicolumn{1}{c|}{with LM} 
& \multicolumn{1}{c}{no LM} & \multicolumn{1}{c|}{with LM} 
& \multicolumn{1}{c}{no LM} & \multicolumn{1}{c|}{with LM} 
& \multicolumn{1}{c}{no LM} & \multicolumn{1}{c}{with LM} \\
\midrule
Greedy CG   & 14.65 & 13.14 & 2250  & 2111  & 24.03 & 22.64  & 1932   & 1802  & 48.68 & 44.88  & 948 & 908 \\
AES        & 14.61 & 12.81 & 71    & 66    & 23.79 & 22.24  & 76     & 71    & \textbf{48.43} & 44.56  & 59  & 53  \\
AES++       & 14.60 & 12.21 & 629   & 522   & 23.78 & 21.75  & 587    & 491   & 48.44 & 42.47  & 450 & 394 \\
ALSD        & \textbf{14.57} & -     & 30    & -     & \textbf{23.77} &    -   & 28     &    -  & 48.58 &    -   & 28  & -   \\
ALSD++      & \textbf{14.57} & \textbf{11.43} & 934   & 788   & \textbf{23.77} & \textbf{20.74}  & 840    & 699   & 48.57 & \textbf{40.70}   & 587 & 532 \\
ALSD++ CG   & \textbf{14.57} & \textbf{11.43} & 1917  & 1780  & \textbf{23.77} & \textbf{20.74}  & 1666   & 1588  & 48.57 & \textbf{40.70}   & 920 & 887 \\
\midrule
\multicolumn{13}{l}{\textbf{RNN-T ($\sim$24k hours)}} \\
\midrule
Greedy CG & \textbf{5.45} & 5.04 & 2268 & 2110   & 10.13 & 10.03 & 1929 & 1805    & 22.87 & 20.53 & 951 & 887 \\
ALSD++ CG & 5.66 & \textbf{4.39} & 1941 & 1798   & \textbf{10.01} & \textbf{9.64}  & 1694 & 1584    & \textbf{22.38} & \textbf{18.80} & 913 & 871 \\
\midrule
\multicolumn{13}{l}{\textbf{TDT ($\sim$24k hours)}} \\
\midrule
Greedy CG & 6.37 & 6.24 & 2605 & 2348 & 10.23 & 10.10   & 2136 & 2001 & 23.42   & 20.82 & 988 &	972 \\
ALSD++ CG & \textbf{5.87} & \textbf{4.43} & 2495 & 2279 & \textbf{10.21} & \textbf{9.79}    & 2045 & 1902 & \textbf{22.95}   & \textbf{19.50} & 977 & 971 \\
\bottomrule
\end{tabular}
}
\label{tab:decoding_comparison}
\vspace{-2em} 
\end{table*}

\subsection{Modification of ALSD and AES}

\textbf{AES:} uses a frame-synchronous decoding strategy, where all hypotheses within the beam are aligned to the same input frame. Decoding progresses sequentially through the input frames, limiting the number of non-token expansions per frame. Prediction and joint network calls for hypotheses in the beam are processed in batch, while the algorithm proceeds sequentially afterward. AES also includes a prefix search, where probabilities of hypotheses with shared prefixes are combined. Using proposed techniques, we implemented fully batched AES: AES++. In AES++, we support only maximum-length prefix search, as shorter prefixes added complexity without improving accuracy.

\textbf{ALSD:} uses a length-synchronous decoding strategy, where all hypotheses within the beam have same token count, including blanks. Prediction and joint network calls for hypotheses in the beam are processed in batch, while the algorithm proceeds sequentially afterward. Prediction network caching is used to avoid redundant evaluations. 
In the original algorithm, the decoding steps are estimated as \( S = T + U_{\text{max}} \), where \( T \) is the input frame count and \( U_{\text{max}} \) is the max output length. Using proposed techniques, we implemented fully batched ALSD: ALSD++. Unlike the original, which forces exactly \( S \) iterations, ALSD++ iterates over \( T \) input frames with a maximum of \( s \) expansions per frame. This prevents hallucinations at the end of outputs. ALSD++ does not use prediction network caching.

Note that the AES algorithm requires more frequent updates to prediction network compared to ALSD, which can be computationally expensive. We encourage readers to refer to our code for further implementation details.

\subsection{Language model fusion with Transducer models}
We apply the proposed beam search algorithms to context biasing via shallow fusion with an external n-gram LM, trained with the same textual representation as the ASR model (same BPE \cite{bpe} vocabulary $V$ is our case). At each step of the beam search, the LM’s probability distribution is combined with the ASR model’s predictions to rescore hypotheses within the beam, assigning higher probabilities to linguistically plausible sequences and reducing errors like misrecognition and syntactic inconsistencies.

\textbf{Scoring of the blank: } Transducer models work in a sequence transduction setting, where the output sequence includes a special blank symbol ($\emptyset$), representing ``next frame transition'', unlike in CTC, where blanks serve as indicators of empty output or separators between tokens. However, standalone LMs, such as N-grams, do not include blank symbol and cannot assign probabilities to blank emissions directly. Thus, we work with two probability distributions:  $p \in \mathbb{R}^{|V|+1}$ from the Transducer model and $p_{\mathrm{LM}} \in \mathbb{R}^{|V|}$ from the LM. Total probability is computed as a weighted sum of both.

In practice, during shallow fusion, the blank symbol's score is typically omitted, and the LM is applied only to non-blank emissions, resulting in combined score of:
\vspace{-2mm}
\begin{equation}
    \ln p_{\text{tot}}[k] =
    \begin{cases}
      \ln p[k] + \lambda \ln p_{\text{LM}}[k], & k \in V \\
      \ln p[\emptyset], & k =\emptyset
    \end{cases} 
    \label{eq:no_score}
\end{equation}
where $\lambda$ is the LM weighting factor. One approach to improving blank scoring modifies the Transducer architecture by introducing separate prediction networks for labels and blanks \cite{article:hat, article:mhat}. Inspired by this, we propose a novel blank scoring method for standard Transducer that adjusts LM probabilities based on ASR blank scores and preserves total probability:
\vspace{-2mm}
\begin{equation}
    \ln p_{\text{tot}}[k] =
    \begin{cases}
      \ln p[k] + \lambda \ln\left((1-p[\emptyset]) p_{\text{LM}}[k] \right),&k \in V \\
      \left(1 + \lambda\right) \ln  p[\emptyset],&k =\emptyset
    \end{cases}
    \label{eq:blank_score}
\end{equation}
In first approach \eqref{eq:no_score} non-blank token scores combine ASR and LM contributions while blank token scores rely solely on ASR, thus increasing the LM weight disproportionately lowers non-blank token probabilities. This imbalance leads to more frequent blank predictions and a higher deletion rate, particularly at higher LM weights. In contrast, the proposed blank scoring method, \eqref{eq:blank_score}, penalizes blank predictions, thus preventing an increase in deletion rates.

\textbf{Early/late pruning:} Distinguished by the point at which ASR and LM scores are combined during beam search, LM fusion has two variations. In early pruning, initially ASR scores are used to select the top $\mathrm{Beam}$ hypotheses, which are then rescored with the LM. This requires only $\mathrm{Beam}$ LM evaluations, which for small beam sizes can be handled with CPU-based sequential LMs. In late pruning, ASR and LM model scores are first combined, and then the $\mathrm{Beam}$ best hypotheses are selected. This increases the LM’s influence but requires LM probabilities over the full vocabulary $V$ $\left(|V|\sim10^3\right)$. CPU-based LMs with sequential processing would significantly slow down decoding, so we explore late pruning using a GPU-accelerated n-gram LM (NGPU-LM)~\cite{ngpulm} that allows batched queries, and causes minimal slowdown.
\section{Experimental Setup}

\textbf{ASR models:} We evaluate low-resource and high-resource setups. In the low-resource scenario, we use a Fast Conformer Transducer Large model with around 115M parameters trained on 960 hours of LibriSpeech. In the high-resource scenario, we use RNN-T~\cite{model:rnnt} and TDT~\cite{model:tdt} of similar architecture and size, trained on 24,000 hours of diverse English speech. All models use a BPE tokenizer with a vocabulary size of 1024.

\textbf{Baseline decodings:} We compare our beam search method with decoding techniques from the NeMo framework \cite{article:nemo}, including the fastest greedy method -- Label-Looping with CUDA-Graphs enabled, and two beam search methods: AES and ALSD. Label-Looping is fully batched, while AES and ALSD support batch processing for the encoder but decode sequentially. NeMo modifies AES by integrating prediction network caching for improved speed.

\textbf{Evaluation datasets:} We report results on three publicly available speech datasets: SPGISpeech \cite{article:spgi} (financial and business domains), SLURP \cite{article:slurp} (single-turn home assistant interactions), and Europarl \cite{article:europarl} (European parliamentary debates). Europarl is also used for 24k training set. For each dataset, we use 6-gram language models built on training set texts for corresponding experiments.

\textbf{N-gram Language models:}
The baseline AES supports N-gram language model fusion, where the fusion strategy corresponds to early pruning without scoring blank tokens \eqref{eq:no_score} in our notation. It employs a CPU-based KenLM N-gram LM \cite{article:kenlm}, which processes each token sequentially. In contrast, our approach uses an NGPU-LM~\cite{ngpulm}, which computes probabilities for the entire vocabulary in parallel on a GPU. Both models have the same underlying ARPA format file, guaranteeing identical LM outputs. Fast batched LM processing enables efficient context biasing during greedy decoding, and we compare our approach with this method. To avoid overfitting, we tune LM fusion parameters using development sets in all experiments.

\textbf{Metrics:} We evaluate decoding accuracy using WER. To assess speed, we report the full system and decoder-only inverse Real-Time Factors (RTFx), computed after a single warm-up run and averaged over three runs. All performance measurements are conducted on a single A5000 GPU using double-precision floating-point numbers.
\begin{table}[t]
    \caption{Full model and decoder-only RTFx(divided by 10) for various batch sizes. \textbf{Model:} RNN-T, TDT. \textbf{Data:} SPGI test. \textbf{Strategy:} Greedy CG (Gr), ALSD++ CG (Bm). \textbf{Beam size:} 6}
    \vspace{-1em} 
    \centering
    \resizebox{\columnwidth}{!}{
    \small
    \begin{tabular}{c|rl|rl|rl|rl}
    \toprule
    & \multicolumn{4}{c}{\textbf{RNNT RTFx$\uparrow$} / $10$} & \multicolumn{4}{|c}{\textbf{TDT RTFx$\uparrow$} / $10$} \\
    \textbf{Batch} & \multicolumn{2}{c}{\textbf{Full}} & \multicolumn{2}{c}{\textbf{Dec.}} & \multicolumn{2}{|c}{\textbf{Full}} & \multicolumn{2}{c}{\textbf{Dec.}} \\
    &   \multicolumn{1}{c}{Gr} & \multicolumn{1}{c}{Bm} &         
        \multicolumn{1}{c}{Gr} & \multicolumn{1}{c}{Bm} & 
        \multicolumn{1}{|c}{Gr} & \multicolumn{1}{c}{Bm} & 
        \multicolumn{1}{c}{Gr} & \multicolumn{1}{c}{Bm} \\
    \midrule
    1   & 29    & 17    & 99    & 30    & 32    & 27    & 145   & 76  \\
    4   & 89    & 61    & 242   & 106   & 105   & 89    & 439   & 246  \\
    8   & 145   & 105   & 388   & 189   & 180   & 151   & 771   & 411  \\
    16  & 190   & 151   & 563   & 308   & 231   & 200   & 1180  & 660  \\
    32  & 228   & 194   & 890   & 525   & 261   & 250   & 1919  & 1029 \\
    \bottomrule
    \end{tabular}%
    }
    \label{tab:comparison}
    \vspace{-1em} 
\end{table}

\begin{table}[t]
    \caption{WER vs RTFx (full model) for multiple beam sizes. \textbf{Model:} RNN-T. \textbf{Data:} SLURP test. \textbf{Strategy:} ALSD, ALSD++. \textbf{CG} - denotes CUDA graphs. \textbf{Batch size:} 32}
    \vspace{-1em} 
    \centering
    \resizebox{\columnwidth}{!}{
    \small
    \begin{tabular}{c|cc|cc|cc}
    \toprule
    \multirow{2}{*}{\textbf{Beam}} & \multicolumn{2}{c|}{\textbf{ALSD}} & \multicolumn{2}{c|}{\textbf{ALSD++ CG}} & \multicolumn{2}{c}{\textbf{ALSD++ CG LM}} \\
    & \multicolumn{1}{c}{\textbf{WER}$\downarrow$} & \multicolumn{1}{c|}{\textbf{RTFx}$\uparrow$} 
    & \multicolumn{1}{c}{\textbf{WER}$\downarrow$} & \multicolumn{1}{c|}{\textbf{RTFx}$\uparrow$} 
    & \multicolumn{1}{c}{\textbf{WER}$\downarrow$} & \multicolumn{1}{c}{\textbf{RTFx}$\uparrow$} \\
    \midrule
    1 & 22.87 & 205 & 22.87 & 983 & 22.87 & 954 \\
    2 & 22.68 & 57 & 22.66 & 938 & 20.91 & 918 \\
    4 & 22.47 & 39 & 22.47 & 925 & 19.21 & 889 \\
    8 & \textbf{22.36} & 22 & \textbf{22.37} & 878 & 18.68 & 826 \\
    12 & 22.37& 13 & 22.38 & 834 & \textbf{18.56} & 774 \\
    \bottomrule
    \end{tabular}}
    \label{tab:speed}
    \vspace{-2em} 
\end{table}

\section{Results}
A summary of our results is shown in Tab. \ref{tab:decoding_comparison}.
Both AES and ALSD offer minor improvements in WER, when decoded without external LM. AES++ and ALSD++ achieve similar accuracy to their baselines while delivering significantly faster performance. ALSD++ outperforms AES++ in terms of speed, due less frequent prediction network updates.

\textbf{Low-resource: } The proposed beam search decoding with an external LM results in a 13.7–22.0\% relative WER improvement across test sets with minimal speed decrease. ALSD++ with an external LM outperforms AES++ in accuracy, likely due to its consistent hypothesis length. Combined with the proposed blank scoring \eqref{eq:blank_score}, this ensures balanced scores within the beam. In contrast, AES++ has varying token counts, which may lead to shorter hypotheses being assigned higher probabilities, making longer hypotheses more likely to be pruned. The blank scoring approach, combined with late pruning, improves overall LM fusion accuracy by 2.2-4.7\% for AES++ and 6.7-10.8\% for ALSD++ compared to baseline AES.

\textbf{CUDA graphs:} The superior speed and accuracy of ALSD++ motivated further optimization through the integration of CUDA Graphs, resulting in our fastest and the most accurate beam search: ALSD++ CG. We then adapted ALSD++ CG for TDT models. For larger batch sizes ALSD++ CG achieves near-greedy full decoding speed, reducing the performance gap to less than 20\% for traditional Transducer models and under 10\% for TDT models.

\begin{figure}[t]
  \centering
    \vspace{-1em}
    \includegraphics[width=0.70\linewidth]{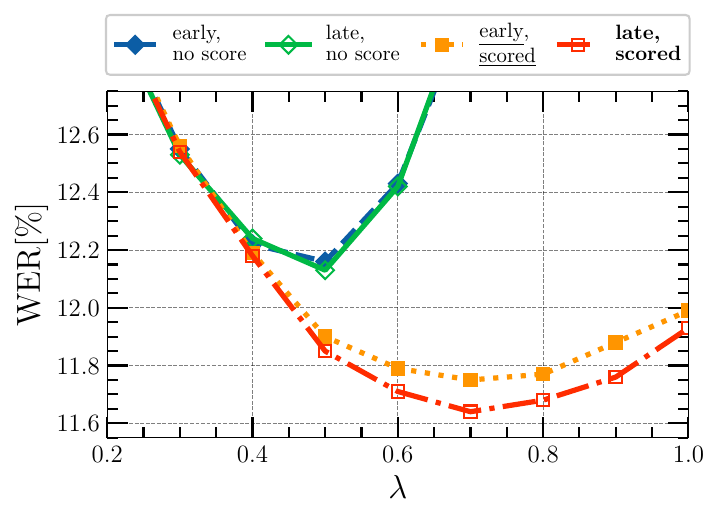}
    \vspace{-10pt}
    \caption{WER vs $\lambda$ (LM weight) for four LM fusion strategies. \textbf{Model:} RNN-T $\left(\sim1k\right)$. \textbf{Data:} SPGI dev. \textbf{Strategy:} ALSD++. \textbf{Beam size:} 6.}
    \label{fig:scoring}
    \vspace{-20pt}
\end{figure}

\textbf{High-resource:} ASR models trained on large datasets can still benefit from external LM fusion when adapting to an "unseen" domain. For RNN-T, it achieves a 17.8–19.4\% improvement on new domains and 4.8\% in-domain. For TDT, the improvement ranges from 11.7–30.5\% on new domains and 4.3\% in-domain. In high-resource scenarios, LM weights tend to be small, and four fusion strategies produce similar results. Additionally, we did not observe improvements from blank scoring in TDT models, as they naturally predict blanks less frequently.

\textbf{Ablation studies: Scaling batch size.} Table~\ref{tab:comparison} contains the full model and the decoder-only RTFx values for multiple batch sizes. For larger batch sizes, the full model's RTFx values for beam search approach those of greedy decoding. However, the decoder-only RTFx values for beam decoding are several times higher ($1.7$–$1.9$ for batch size 32). At larger batch sizes, the performance bottleneck shifts to the encoder, and the decoder part is relatively lightweight.

\textbf{Ablation studies: Scaling beam size.} Tab.~\ref{tab:speed} contains the full model RTFx values for multiple beam sizes. The baseline ALSD (and AES) algorithms slow down significantly as beam size increases, whereas ALSD++ (and AES++) leverage batched processing to keep the slowdown minimal. Larger beams in LM fusion improve decoding accuracy, as hypothesis search space becomes wider.

\textbf{Ablation studies: Blank scoring.} We evaluated four LM fusion strategies: early and late pruning, each combined with one of two blank scoring approaches \eqref{eq:no_score}, \eqref{eq:blank_score}. In low-resource settings, where larger LM values are beneficial, late pruning with blank scoring consistently improves accuracy, benefiting both ALSD++ and AES++. Figure~\ref{fig:scoring} shows WER as a function of LM weights in a low-resource scenario. The proposed blank scoring method \eqref{eq:blank_score} prevents deletion rate increase at higher LM weights, significantly enhancing recognition. Late pruning offers a slight additional accuracy boost.

\section{Conclusion}
Beam search decoding has traditionally been a bottleneck for Transducer models, significantly slowing down inference compared to greedy decoding. In this paper, we introduced a universal beam search acceleration method that enables two novel fast decoding algorithms, ALSD++ and AES++. Our approach integrates batched processing, a tree-based hypothesis structure, a novel blank scoring technique for LM fusion, and CUDA graph execution for optimized GPU performance. By leveraging these innovations, we reduce the speed gap between beam search and greedy decoding to just 10–20\% for the whole system, while achieving a relative WER improvement of 14–30\% in key scenarios. All proposed algorithms are available as open source.

\ifinterspeechfinal
\else
     
\fi

\newpage
\bibliographystyle{IEEEtran}
\bibliography{mybib}

\begin{thebibliography}{10}
\providecommand{\url}[1]{#1}
\csname url@samestyle\endcsname
\providecommand{\newblock}{\relax}
\providecommand{\bibinfo}[2]{#2}
\providecommand{\BIBentrySTDinterwordspacing}{\spaceskip=0pt\relax}
\providecommand{\BIBentryALTinterwordstretchfactor}{4}
\providecommand{\BIBentryALTinterwordspacing}{\spaceskip=\fontdimen2\font plus
\BIBentryALTinterwordstretchfactor\fontdimen3\font minus \fontdimen4\font\relax}
\providecommand{\BIBforeignlanguage}[2]{{%
\expandafter\ifx\csname l@#1\endcsname\relax
\typeout{** WARNING: IEEEtran.bst: No hyphenation pattern has been}%
\typeout{** loaded for the language `#1'. Using the pattern for}%
\typeout{** the default language instead.}%
\else
\language=\csname l@#1\endcsname
\fi
#2}}
\providecommand{\BIBdecl}{\relax}
\BIBdecl

\bibitem{graves_transducer}
A.~Graves, ``Sequence transduction with recurrent neural networks,'' in \emph{ICML}, 2012.

\bibitem{ShallowFusionEC}
D.~Zhao, T.~N. Sainath, D.~Rybach, P.~Rondon, D.~Bhatia, B.~Li, and R.~Pang, ``Shallow-fusion end-to-end contextual biasing,'' in \emph{Interspeech}, 2019.

\bibitem{Andrusenko2024FastCF}
A.~Andrusenko, A.~Laptev, V.~Bataev, V.~Lavrukhin, and B.~Ginsburg, ``Fast context-biasing for {CTC} and {Transducer ASR} models with {CTC-based} word spotter,'' \emph{Interspeech}, 2024.

\bibitem{Nolden2011AcousticLF}
D.~Nolden, R.~Schl{\"u}ter, and H.~Ney, ``Acoustic look-ahead for more efficient decoding in {LVCSR},'' in \emph{Interspeech}, 2011.

\bibitem{Jain2019RNNTFL}
M.~Jain, K.~Schubert, J.~Mahadeokar \emph{et~al.}, ``{RNN-T} for latency controlled {ASR} with improved beam search,'' \emph{ArXiv/1911.01629}, 2019.

\bibitem{alsd}
G.~Saon, Z.~T{\"u}ske, and K.~Audhkhasi, ``{Alignment-Length Synchronous Decoding for RNN Transducer},'' \emph{ICASSP}, 2020.

\bibitem{maes}
J.~Kim, Y.~Lee, and E.~Kim, ``Accelerating {RNN Transducer} inference via adaptive expansion search,'' \emph{IEEE Signal Processing Letters}, 2020.

\bibitem{Keren2023ATB}
G.~Keren, ``A token-wise beam search algorithm for {RNN-T},'' \emph{ASRU}, 2023.

\bibitem{kang2023fastparallel}
W.~Kang, L.~Guo, F.~Kuang, L.~Lin, M.~Luo, Z.~Yao, X.~Yang, P.~{\.Z}elasko, and D.~Povey, ``Fast and parallel decoding for transducer,'' in \emph{ICASSP 2023-2023 IEEE International Conference on Acoustics, Speech and Signal Processing (ICASSP)}.\hskip 1em plus 0.5em minus 0.4em\relax IEEE, 2023, pp. 1--5.

\bibitem{stateless}
M.~R. Ghodsi, X.~Liu, J.~A. Apfel, R.~Cabrera, and E.~Weinstein, ``{RNN-Transducer} with stateless prediction network,'' \emph{ICASSP}, 2020.

\bibitem{Wang2022AcceleratingRT}
Y.~Wang, Z.~Chen, C.~yong Zheng, Y.~Zhang, W.~Han, and P.~Haghani, ``Accelerating {RNN-T} training and inference using {CTC} guidance,'' \emph{ICASSP}, 2022.

\bibitem{tdt}
H.~Xu, F.~Jia, S.~Majumdar, H.~Huang, S.~Watanabe, and B.~Ginsburg, ``Efficient sequence transduction by jointly predicting tokens and durations,'' \emph{ICML}, 2023.

\bibitem{Mahadeokar2022StreamingPT}
J.~Mahadeokar, Y.~Shi, K.~Li, D.~Le, J.~Zhu, V.~Chandra, O.~Kalinli, and M.~L. Seltzer, ``Streaming parallel transducer beam search with fast-slow cascaded encoders,'' in \emph{Interspeech}, 2022.

\bibitem{article:labellooping}
V.~Bataev, H.~Xu, D.~Galvez, V.~Lavrukhin, and B.~Ginsburg, ``Label-looping: Highly efficient decoding for transducers,'' in \emph{SLT}, 2024.

\bibitem{Galvez2024SpeedOL}
D.~Galvez, V.~Bataev, H.~Xu, and T.~Kaldewey, ``Speed of light exact greedy decoding for {RNN-T} speech recognition models on {GPU},'' \emph{Interspeech}, 2024.

\bibitem{article:nemo}
O.~Kuchaiev, J.~Li, H.~Nguyen, O.~Hrinchuk, R.~Leary, B.~Ginsburg, S.~Kriman, S.~Beliaev, V.~Lavrukhin, J.~Cook, P.~Castonguay, M.~Popova, J.~Huang, and J.~M. Cohen, ``Nemo: a toolkit for building ai applications using neural modules,'' 2019.

\bibitem{article:espnet}
S.~Watanabe, T.~Hori, S.~Karita, T.~Hayashi, J.~Nishitoba, Y.~Unno, N.~{Enrique Yalta Soplin}, J.~Heymann, M.~Wiesner, N.~Chen, A.~Renduchintala, and T.~Ochiai, ``{ESPnet}: End-to-end speech processing toolkit,'' in \emph{Interspeech}, 2018.

\bibitem{graves_ctc}
A.~Graves, S.~Fern{\'a}ndez, F.~J. Gomez, and J.~Schmidhuber, ``Connectionist temporal classification: labelling unsegmented sequence data with recurrent neural networks,'' \emph{ICML}, 2006.

\bibitem{bpe}
R.~Sennrich, B.~Haddow, and A.~Birch, ``Neural machine translation of rare words with subword units,'' in \emph{Proceedings of the 54th Annual Meeting of the Association for Computational Linguistics}, 2016.

\bibitem{article:hat}
E.~Variani, D.~Rybach, C.~Allauzen, and M.~Riley, ``Hybrid autoregressive transducer (hat),'' \emph{ICASSP}, 2020.

\bibitem{article:mhat}
Z.~Meng, T.~Chen, R.~Prabhavalkar, and other, ``Modular hybrid autoregressive transducer,'' \emph{SLT}, 2022.

\bibitem{ngpulm}
V.~Bataev, A.~Andrusenko, L.~Grigoryan, A.~Laptev, V.~Lavrukhin, and B.~Ginsburg, ``{NGPU-LM}: {GPU}-{A}ccelerated {N}-{G}ram {L}anguage {M}odel for context-biasing in greedy {ASR} decoding,'' \emph{ArXiv/2505.22857}, 2025.

\bibitem{model:rnnt}
\BIBentryALTinterwordspacing
NVIDIA, ``{STT En FastConformer Transducer Large},'' 2020. [Online]. Available: \url{https://hf.co/nvidia/stt_en_fastconformer_transducer_large}
\BIBentrySTDinterwordspacing

\bibitem{model:tdt}
\BIBentryALTinterwordspacing
{NVIDIA}, ``{STT En FastConformer TDT Large},'' 2020. [Online]. Available: \url{https://hf.co/nvidia/stt_en_fastconformer_tdt_large}
\BIBentrySTDinterwordspacing

\bibitem{article:spgi}
P.~K. {O'Neill}, V.~{Lavrukhin}, S.~{Majumdar}, V.~{Noroozi}, Y.~{Zhang}, O.~{Kuchaiev}, J.~{Balam}, Y.~{Dovzhenko}, K.~{Freyberg}, M.~D. {Shulman}, B.~{Ginsburg}, S.~{Watanabe}, and G.~{Kucsko}, ``{{SPGISpeech}: 5,000 hours of transcribed financial audio for fully formatted end-to-end speech recognition},'' \emph{arXiv:2104.02014}, 2021.

\bibitem{article:slurp}
E.~Bastianelli, A.~Vanzo, P.~Swietojanski, and V.~Rieser, ``{SLURP: A Spoken Language Understanding Resource Package},'' in \emph{{EMNLP}}, 2020.

\bibitem{article:europarl}
G.~V. Garcés Díaz-Munío, J.~A. Silvestre-Cerdà, J.~Jorge, A.~Giménez, J.~Iranzo-Sánchez, P.~Baquero-Arnal, N.~Roselló, A.~P.-G. de~Martos, J.~Civera, A.~Sanchis, and A.~Juan, ``{Europarl-ASR: A Large Corpus of Parliamentary Debates for Streaming ASR Benchmarking and Speech Data Filtering/Verbatimization},'' in \emph{Interspeech}, 2021.

\bibitem{article:kenlm}
K.~Heafield, ``{K}en{LM}: Faster and smaller language model queries,'' in \emph{Proc. of the Sixth Workshop on Statistical Machine Translation}, 2011.

\end{thebibliography}

\end{document}